\documentclass{PoS}

\usepackage{amsmath,graphicx,verbatim}
\usepackage{epstopdf}
\usepackage[utf8x]{inputenc}
\usepackage{amsmath,graphicx}
\usepackage{simplewick}
\usepackage{datetime}
\usepackage{here}
\usepackage{setspace}

\newcommand{\Ga}{\Gamma}

\newcommand{\nn}{\nonumber}

\newcommand{\be}{\begin{equation}}
\newcommand{\ee}{\end{equation}}
\newcommand{\bea}{\begin{eqnarray}}
\newcommand{\eea}{\end{eqnarray}}
\newcommand{\balign}{\begin{align}}
\newcommand{\ealign}{\end{align}}
\newcommand{\as}{\alpha_s}

\newcommand{\cd}{\cdot}

\newcommand{\bg}{\begin{gather}}
\newcommand{\foma}{\end{gather}}

\newcommand{\noopsort}[1]{}

\newcommand{\vecb}[1]{\mbox{\boldmath $#1$}}
\newcommand{\vecbe}[1]{\mbox{\boldmath ${\scriptstyle #1}$}}

\def\L{\Lambda}

\def\z{\zeta}

\def\<{\langle}
\def\>{\rangle}

\def\b{\beta}
\def\g{\gamma}  \def\G{\Ga}
  \def\D{\Delta}
\def\l{\lambda}   \def\L{\Lambda}
\def\s{\sigma}

\def\m{\mu}
\def\n{\nu}

\def\z{\zeta}

\def\({\left(}
\def\[{\left[}
\def\){\right)}
\def\]{\right]}

\def\ln{\hbox{ln}}

\def\le{\left }
\def\ri{\right}

\def\lqcd{\L_{\rm QCD}}

\newcommand{\ben}{\begin{eqnarray}}
\newcommand{\een}{\end{eqnarray}}

\newcommand{\bef}{\begin{figure}[htb]\centering}
\newcommand{\eef}{\end{figure}}

\newcommand{\eq}[1]{Eq.~\eqref{#1}}

\title{Transverse Momentum Dependent (Un)polarized Gluon Distributions in Higgs Production}

\ShortTitle{Transverse Momentum Dependent (Un)polarized Gluon Distributions in Higgs Production}

\author{\speaker{Tomas KASEMETS}%
\\
       Nikhef Theory Group and VU University Amsterdam, \\De Boelelaan 1081, NL-1081 HV Amsterdam, the Netherlands\\
       E-mail: \email{kasemets@nikhef.nl}}

\abstract{We combine soft and collinear matrix elements to define transverse momentum dependent parton distributions (TMDs) for gluons, free from rapidity divergences. We establish a factorization theorem at next-to-leading order for the Higgs boson transverse momentum ($q_T$) spectrum, and use it to derive evolution equations for gluon TMDs. The evolution for all gluon TMDs is driven by a universal kernel, i.e. the same for all polarizations. In the region of intermediate $q_T$ we match the unpolarized, helicity and linearly polarized gluon distributions onto PDFs. We calculate the resummed Higgs transverse momentum distribution at NNLL, including the contribution of the linearly polarized gluons, and investigate the impact of non-perturbative models for the transverse distribution of gluons inside the proton.}

\FullConference{XXIII International Workshop on Deep-Inelastic Scattering\\
		 27 April - May 1 2015\\
		 Dallas, Texas}

\begin{document}

\section{Factorization theorem and evolution equations for gluon TMDs}
\label{sec:intro}
Much effort has been devoted to properly describe the transverse momentum dependent parton distributions (TMDs), in order to avoid undesired features preventing them from representing physical hadronic quantities -- such as spurious rapidity divergences. Recently, improved definitions were introduced, regulating and canceling these divergencies and thus removing the bad features from the quark TMDs  \cite{Collins:2011zzd, GarciaEchevarria:2011rb,Echevarria:2012js}. We extend these methods and definitions to properly define also the gluon TMDs in \cite{Echevarria:2015uaa}. 

In order to properly define all the (un)polarized gluon TMDs we derive a factorization theorem, separating the physics at different scales, for the transverse momentum spectrum in Higgs boson production through gluon fusion, and explicitly verify it at next-to-leading order (NLO). The factorization formula is used to derive the evolution equations  for all the gluon TMDs, which are driven by a single universal kernel. For intermediate transverse momenta of the Higgs boson, the TMDs can be re-factorized and matched onto the canonical PDFs. Combined with the evolution equations this allows the resummation of large $q_T$ logarithms. The known perturbative ingredients allow us to perform the resummation of the Higgs $q_T$-spectrum at next-to-next-to-leading-logarithmic  (NNLL) accuracy. We supplement the resummation with modeling of the non-perturbative part of the TMDs and study the impact of the non-perturbative physics on the Higgs transverse momentum distribution. Thus we provide a general framework to deal with gluon TMDs in different processes and account for their perturbative and non-perturbative contributions. 

The $q_T$ distribution of the Higgs boson has received much attention, both in the context of perturbative QCD (see for example, \cite{Sun:2011iw,Catani:2011kr}) and soft-collinear effective theory (SCET) (see, e.g., \cite{Chiu:2012ir,Mantry:2009qz,Becher:2012yn,Neill:2015roa}). TMD gluon correlators were also considered in~\cite{Ji:2005nu,Zhu:2013yxa}. However none of the previous works paid attention to the cancellation of rapidity divergences in a proper definition of gluon TMDPDFs.
We want to emphasize, that our point of focus is on the gluon TMDs, and we do not aim to obtain the most precise predictions for the Higgs $q_T$ spectra.

An interesting feature of gluon TMDs, is that there are two different distributions even in an unpolarized proton, one for unpolarized and one for linearly polarized gluons \cite{Mulders:2000sh,Catani:2010pd}.
In \cite{Boer:2014tka} the authors quantified their contribution in the context of the TMD formalism, both for Higgs boson and $C$-even scalar quarkonium ($\chi_{c0}$ and $\chi_{b0}$) production. We extend their efforts by implementing the currently known perturbative ingredients to the full extent to perform the resummation at NNLL accuracy, providing more accurate predictions and investigating their uncertainty. 

We make use of the effective theory point of view, in which the factorization theorem is derived as a stepwise matching, 
$\text{QCD}_{n_f=6} \rightarrow \text{QCD}_{n_f=5} \rightarrow \text{SCET}_{q_T} \rightarrow \text{SCET}_{\Lambda_{\text{QCD}}} \, .$
In a first step, we integrate out the top-quark, leading to an effective theory for QCD with $n_f=5$ active flavors and an effective ggH vertex. Subsequently the hard modes of the gluon and quark fields, with off-shellness at the scale of the Higgs boson mass, are integrated out and we are left with collinear an soft fields described by SCET$_{q_T}$. In a final step, valid only for perturbatively large transverse momenta $\Lambda_{QCD} << q_T << Q$, the TMDs are matched onto PDFs, with an effective description in SCET$_{\Lambda_{\text{QCD}}}$. In each step, we obtain a matching coefficient, correcting for the differences in the ultraviolet region of the theories.

The cross section can be expressed in terms of collinear and soft matrix elements, which  contain rapidity divergencies and thus are ill-defined, but which can be carefully combined in order to cancel them and give well defined gluon TMDs, see \cite{Echevarria:2015uaa} for a detailed description. There are different methods to deal with the rapidity divergencies by combining the soft and collinear matrix elements. One can use a rapidity regulator, such as the $\D$ regulator \cite{GarciaEchevarria:2011rb}, the rapidity regulator introduced in \cite{Chiu:2011qc} or alternatively combine the integrands in order to cancel the divergencies \cite{Collins:2011zzd}. We want to emphasize that the proper definition of the TMDs does not depend on the choice of the regulator used \cite{Collins:2012uy}.

In terms of the gluon TMDs, ${\tilde G}_{g/A}$, the cross section for Higgs $q_T$ distribution takes the form
\begin{align}\label{eq:factth}
\frac{d\sigma}{dy\,d^2q_\perp} &= 
2\s_0(\mu)\,C_t^2(m_t^2,\mu) H(m_H^2,\m)\,
\frac{1}{(2\pi)^2} \int d^2y_\perp\,e^{i\vecbe q_\perp\cd\vecbe y_\perp} 
\nn\\
&
\times
{\tilde G}_{g/A}^{\m\n}(x_A,\vecb y_\perp,S_A;\z_A,\m)\,
{\tilde G}_{g/B\,\m\n}(x_B,\vecb y_\perp,S_B;\z_B,\m)
+{\cal O}(q_T/m_H)
\,.
\end{align}
$C_t$ and  $H(m_H^2,\m)=|C_H(-q^2,\m)|^2$ are the coefficients in the two first matchings in the derivation of the factorization theorem. $x_{A,B}=\sqrt{\tau}\,e^{\pm y}$, $\tau=(m_H^2+q_T^2)/s$ and $y$ is the rapidity of the produced Higgs boson. The Born-level cross section is
$\s_0(\mu) = m_H^2\,\as^2(\mu)/(72\pi (N_c^2-1) s v^2) $
 given in terms of the Higgs boson mass $m_H$, the strong coupling $\as$, the square of the proton CM energy $s$ and the vacuum expectation value $v$. Note that the gluon TMDs depend on two scales, the factorization scale $\mu$ and an energy scale $\zeta$ -- related to the separation of the two TMDs in rapidity. The evolution of the gluon TMDs in the renormalization scale $\m$ is driven by the anomalous dimensions,
\begin{align}
\frac{d}{d\ln\m}\ln\tilde{G}^{[pol]}_{g/A}(x_A,\vecb b_\perp,S_A;\z_A,\m) & =
-\G_{\rm cusp}^A(\as(\m))\ln\frac{\z_A}{\m^2} - \g^g(\as(\m))  - \g^t(\as(\m)) - \frac{\b(\as(\m))}{\as(\m)}
\,.
\end{align}
While the evolution in the rapidity scale $\z$ 
\begin{align}\label{eq:evolzeta}
\frac{d}{d\ln\z_A}\ln\tilde{G}^{[pol]}_{g/A}(x_A,\vecb b_\perp,S_A;\z_A,\m) &=
- D_g(b_T;\m)
\,,
\end{align}
is controlled by a function $D_g$ containing both non-perturbative and perturbative information. Thus part of the evolution equations for the TMDs has to be modeled and/or measured. Combining the evolution in the two scales the gluon TMDs evolve as
\begin{align}
{\tilde G}^{[pol]}_{g/A}(x_A,\vecb b_\perp,S_A;\z_{A,f},\m_f) &=
{\tilde G}^{[pol]}_{g/A}(x_A,\vecb b_\perp,S_A;\z_{A,i},\m_i)\,
{\tilde R}^g\le(b_T;\z_{A,i},\m_i,\z_{A,f},\m_f\ri)
\,,
\end{align}
with the evolution kernel
\begin{align}\label{eq:evolkernel}
{\tilde R}^g\big(b_T;\z_{A,i},\m_i,\z_{A,f},\m_f\big) &=
\exp\le\{
\int_{\m_i}^{\m_f} \frac{d\bar\m}{\bar\m}\, 
\g_G\le(\as(\bar\m),\ln\frac{\z_{A,f}}{\bar\m^2} \ri)
\ri\}
\le( \frac{\z_{A,f}}{\z_{A,i}} \ri)^{-D_g\le(b_T;\m_i\ri)}
\,.
\end{align}

\section{Re-factorization of TMDs and $q_T$-resummation}
The gluon TMDs for general polarization can be decomposed into different functions for different proton and gluon polarizations, see e.g. \cite{Mulders:2000sh}. For intermediate transverse momenta $\Lambda_{QCD}^2 << q_T^2 << Q^2$ the gluon TMDs can be matched onto their collinear analogues. There are three functions, $f_1^g$ describing unpolarized gluons in an unpolarized proton, $h_1^{\perp g}$ describing linearly polarized gluons in an unpolarized proton and $g_{1L}^{g}$  describing longitudinally polarized gluons in a longitudinally polarized proton, which can be perturbatively generated by the canonical (leading-twist) PDFs. We calculate the matching coefficients at NLO for the all three functions. 

Depending on the TMD considered, the collinear functions which describe its perturbative, small-$b_T$, region will be different, and also the relevant Wilson coefficients. The operator product expansion for the three functions are given by 
\begin{align}\label{eq:opeunpol}
{\tilde f}_{1}^{g/A}(x_A,b_T;\z_A,\m) &=
\sum_{j=q,\bar q, g} \int_{x_A}^{1}\frac{d\bar x}{\bar x}
{\tilde C}_{g/ j}^{f}(\bar x,b_T;\z_A,\m)\,
f_{j/A}(x_A/\bar x;\m)
+{\cal O}(b_T\lqcd)
\,,\nn\\
{\tilde h}_{1}^{\perp g/A\,(2)}(x_A,b_T;\z_A,\m) &=
\sum_{j=q,\bar q, g} \int_{x_A}^{1}\frac{d\bar x}{\bar x}
{\tilde C}_{g/ j}^{h}(\bar x,b_T;\z_A,\m)\,
f_{j/A}(x_A/\bar x;\m)
+{\cal O}(b_T\lqcd)
\,,\nn\\
{\tilde g}_{1L}^{g/A}(x_A,b_T;\z_A,\m) &=
\sum_{j=q,\bar q,g} \int_{x_A}^{1}\frac{d\bar x}{\bar x}
{\tilde C}_{g/ j}^{g}(\bar x,b_T;\z_A,\m)\,
g_{j/A}(x_A/\bar x;\m)
+{\cal O}(b_T\lqcd)
\,,
\end{align}
where the matching coefficients at NLO are given in \cite{Echevarria:2015uaa}. Note that the TMDs for the unpolarized gluons and the linearly polarized gluons are both matched onto the same PDF, but the first non-zero order of the matching coefficient for the linearly polarized gluons is one order higher in $\alpha_s$. The gluon helicity TMD is matched onto the helicity PDF.

We can choose to set the resummation scale either in impact parameter space or in momentum space. In these proceedings we only present results with the resummation performed in impact parameter space, see \cite{Echevarria:2015uaa} for a more detailed discussion . The resummed TMDs in impact parameter space can be written as
\begin{align}
{\tilde F}_{g/A}^{Pert}(x_A,b_T;\z_A,\m) &=
\exp\le\{
\int_{\m_0}^{\m} \frac{d\bar\m}{\bar\m}
\g_G\le(\as(\bar\m),\ln\frac{\z_A}{\bar\m^2} \ri)\ri\}\,
\le(\frac{\z_A}{\z_0}\ri)^{-D_g(b_T;\m_0)}
\nn\\
&\times
\sum_{j=q,\bar q, g}
{\tilde C}_{g/j}(x_A,b_T;\z_0,\m_0)\otimes
f_{j/A}(x_A;\m_0)\,
\,,
\end{align}
where $\z_0\sim \m_b^2$ and $\m_0\sim \m_b=2e^{-\gamma_E}/b_T$.
The superscript $Pert$ signifies that it is only the perturbative part of the TMDs~\footnote{We refer to the perturbative or non-perturbative nature of the transverse momentum (or impact parameter) dependence, leaving aside the non-perturbative PDFs.}, valid at small $b_T<<1/\Lambda_{QCD}$.
\section{Higgs $q_T$ spectrum}
For large $b_T$ we need to supplement the perturbative expression with a model with parameters that can be extracted from experimental data.
We implement a smooth cutoff that freezes the perturbative contribution towards large $b_T$,
\begin{align}\label{eq:nonp-bhat}
{\tilde F}_{g/A}(x_A,b_T;\z_A,\m) &= 
{\tilde F}_{g/A}^{Pert}(x_A,{\hat b}_T;\z_A,\m)\,
{\tilde F}_{}^{NP}(x_A,b_T;\z_A)
\,,
\end{align}
with the cutoff prescription $\hat b_T(b_T) = b_c\sqrt{( 1 - e^{-b_T^2/b_c^2} )}$ and $b_c$ determining the separation between the perturbative and non-perturbative regions. We parametrize the non-perturbative piece for the two TMDs contributing to Higgs production as
\begin{align}
{\tilde F_{j/A}}^{f,NP} =
\exp\le[-b_T^2(\l_f+\l_Q \ln(Q^2/Q_0^2)) \ri]
\,,\quad
{\tilde F_{j/A}}^{h,NP}=
\exp\le[-b_T^2(\l_h+\l_Q \ln(Q^2/Q_0^2)) \ri]
\,,
\end{align}
where $Q_0 = 1~{\rm GeV}$. $\l_Q$ is the same for both functions, since the evolution is universal among all (un)polarized TMDs, and therefore, their scale-dependence is the same. The non-perturbative model ${\tilde F}^{NP}$ should be $1$ for $b_T=0$ not to modify the perturbative result and plays an increasingly important role as we increase $b_T$.
\begin{figure}[t]
\begin{center}
\includegraphics[width=0.32\textwidth]{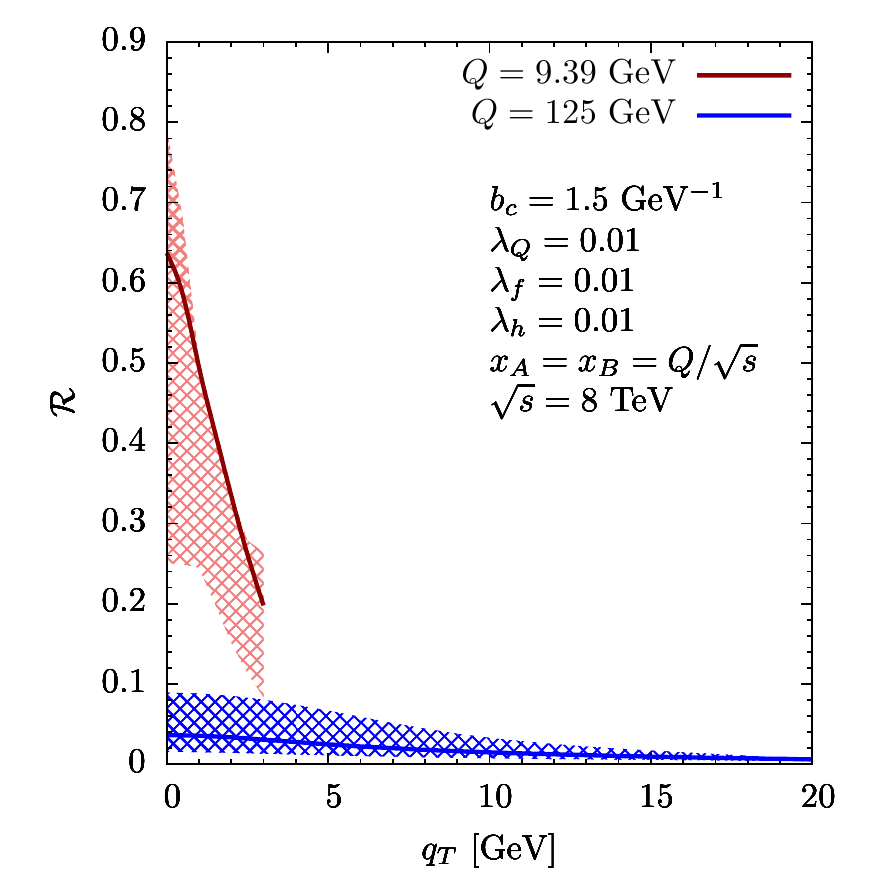}
\quad
\includegraphics[width=0.32\textwidth]{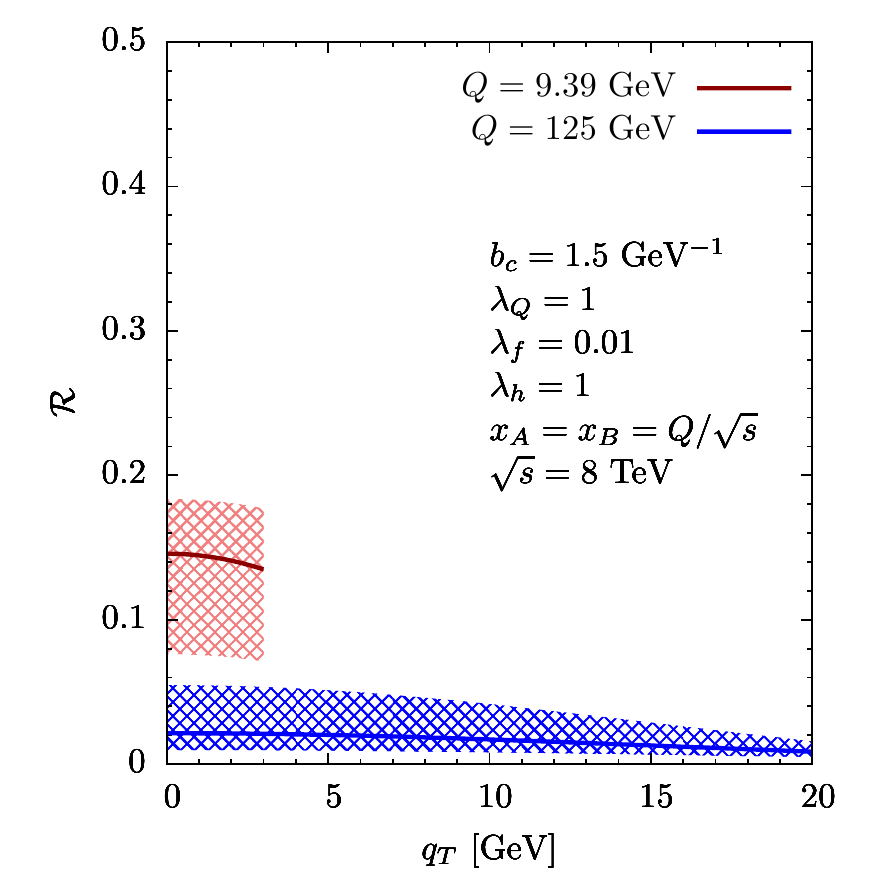}
\end{center}
\caption{\it
Ratio ${\cal R}$ for different values of the non-perturbative parameters $\l_{f(h)}$ and $\l_Q$, at the scales for Higgs boson and $\eta_{b}$ production. 
The curves are calculated at NNLL accuracy and for $\sqrt{s}=8$ TeV.
}
\label{fig:ratio}
\end{figure}
The contribution of unpolarized and/or linearly polarized gluon distributions in unpolarized hadron-hadron collisions depends on the process under study and has been discussed in several works, see for example ~\cite{Boer:2011kf,Boer:2013fca,Pisano:2013cya,Dunnen:2014eta,Boer:2014tka}. In the cases we consider, Higgs boson as well as $\eta_b$ production, both unpolarized and linearly polarized distributions play a role, and we investigate their relative contribution to the cross section.
We use our results to quantify the contribution of linearly polarized gluons, considering the following ratio
\begin{align}
{\cal R}(x_A,x_B,q_T;Q) &=
\frac{\int d^2\vecb b_T\,e^{-i\vecbe q_T\cd \vecbe b_T}
{\tilde h}_{1}^{\perp g/A(2)}(x_A,b_T;Q^2,Q)\,
{\tilde h}_{1}^{\perp g/B(2)}(x_B,b_T;Q^2,Q)}
{\int d^2\vecb b_T\,e^{-i\vecbe q_T\cd \vecbe b_T}
{\tilde f}_{1}^{g/A}(x_A,b_T;Q^2,Q)\,{\tilde f}_{1}^{g/B}(x_B,b_T;Q^2,Q)}
\,,
\end{align}
which determine the relative contribution from linearly polarized and unpolarized gluons to the cross section, for both Higgs boson and $C$-even pseudoscalar bottonium production. 

Fig.~\ref{fig:ratio} shows our results for the ratio ${\cal R}$ at the scales for the transverse momentum distributions of Higgs boson and $\eta_{b}$, all at NNLL accuracy. The bands are obtained by independently varying the scales $\z_0$ and $\m_0$ around their default value by a factor of 2, and plotting the maximum uncertainty for each point in $q_T$. In order to estimate the impact on the ratio of the different non-perturbative parameters, we have chosen several values in a sensible range and selected some combinations in limiting cases. The outcome of the numerical study is clear: the lower the scale the more contribution we have from linearly polarized gluons, although this contribution depends on the value of the non-perturbative parameters.
At the Higgs boson scale the effect of linearly polarized gluons is small, around 1-9\%, making it harder to extract their non-perturbative parameters from experimental data.
At lower scales, as in the production of $\eta_{b}$, their role is enhanced, from 10\% up to 70\%, and thus experimental data can better determine them.
It seems plausible that their non-perturbative parameters can be fixed in the near future by properly combining experimental data for different experiments and at different scales. The framework introduced in this paper, with the proper definition of gluon TMDs and their QCD evolution, will be crucial in order to consistently address different processes in terms of the same hadronic quantities and properly extract their non-perturbative parameters.

After analyzing the contribution of linearly polarized gluons for $\eta_b$ and Higgs boson production in unpolarized hadron-hadron collisions, we turn to the Higgs boson transverse momentum distribution at the LHC. The cross section for this process, given in \eq{eq:factth}, for unpolarized protons is
\begin{align}\label{eq:higgsdistribution}
\frac{d\sigma}{dy\,d^2q_\perp} &= 
2\s_0(\mu)\,C_t^2(m_t^2,\mu) H(m_H^2,\m)\,
\frac{1}{(2\pi)^2} \int d^2y_\perp\,e^{i\vecbe q_\perp\cd\vecbe y_\perp} 
\nn\\
&
\times
\frac{1}{2}\le[
{\tilde f}_1^{g/A}(x_A,b_T;\z_A,\m)\,
{\tilde f}_1^{g/B}(x_B,b_T;\z_B,\m) \ri.
\nn\\
&
+ \le.
{\tilde h}_1^{\perp\, g/A (2)}(x_A,b_T;\z_A,\m)\,
{\tilde h}_1^{\perp\, g/B (2)}(x_B,b_T;\z_B,\m)
\ri]
+
{\cal O}(q_T/m_H)
\,.
\end{align}
The evolution kernel suppresses the TMDs at large $b_T$, and this effect increases with the hard scale $Q$ \cite{Echevarria:2012pw,Qiu:2000hf}. Therefore, the larger $Q$ the less sensitive the resummed expression will be to the non-perturbative contributions.
\begin{figure}[t]
\begin{center}
\includegraphics[width=0.32\textwidth]{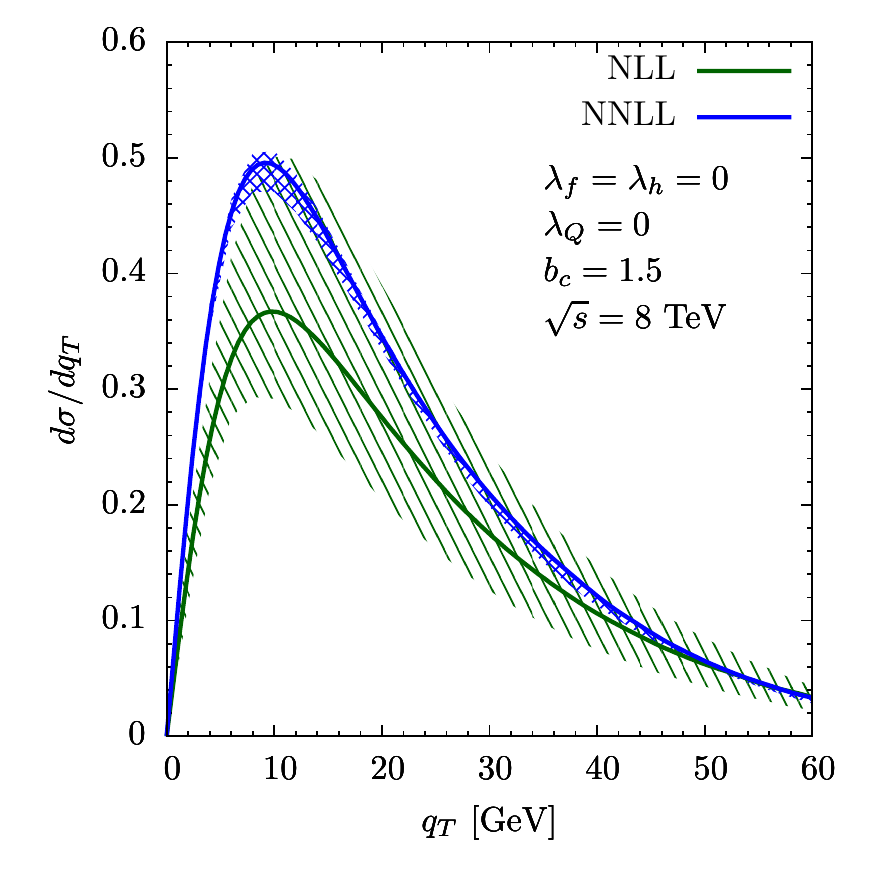}
\includegraphics[width=0.32\textwidth]{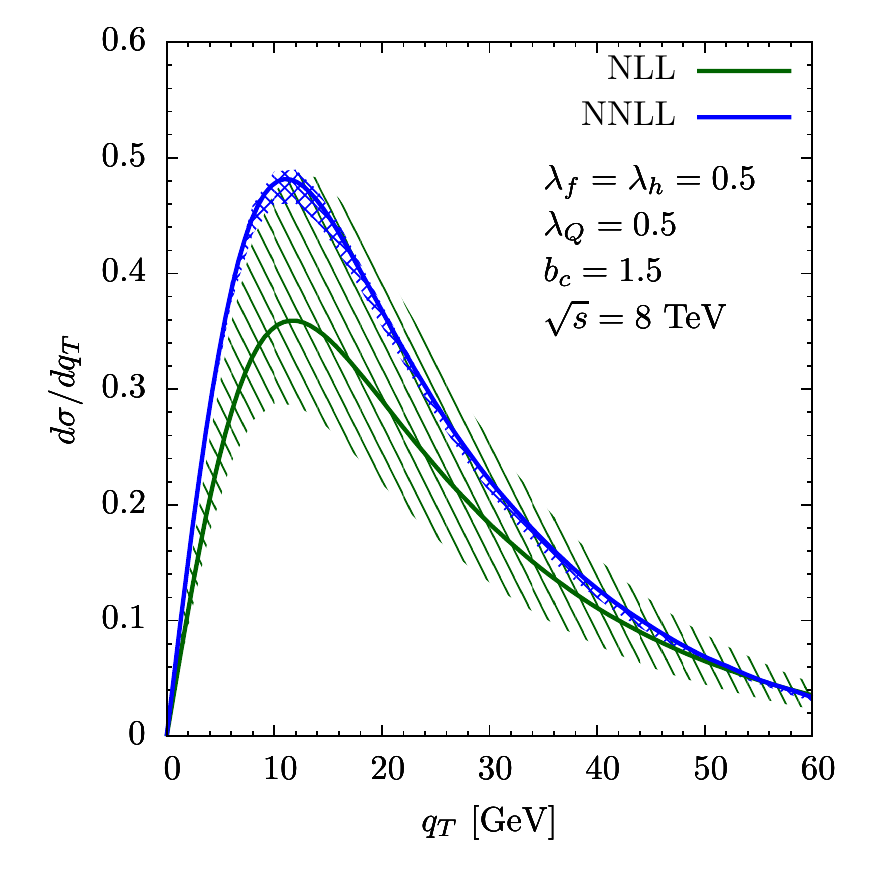}
\includegraphics[width=0.32\textwidth]{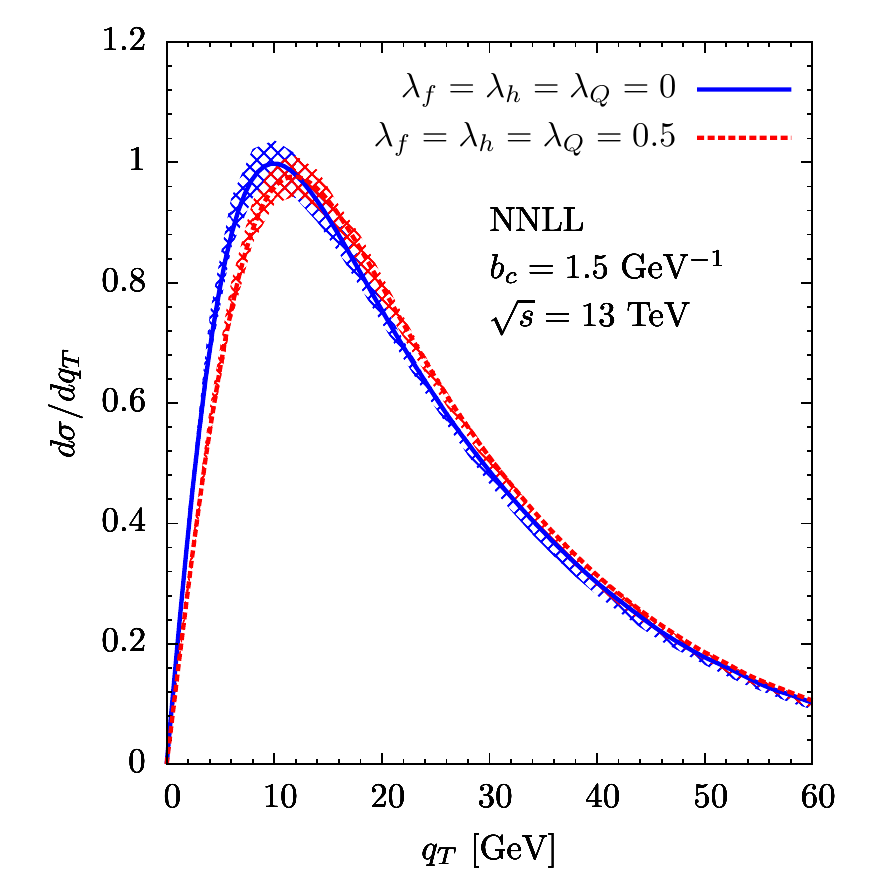}
\end{center}
\caption{\it
Cross section at $\sqrt{s} = 8$~TeV (left and middle) and $13$~TeV for different values of the non-perturbative parameters $\l_{f,h,Q}$, with $\m_0\sim \m_b$ an NLL and NNLL.}
\label{fig:xsection2}
\end{figure}
The first two panels of Figure~\ref{fig:xsection2} show the Higgs boson transverse momentum distribution at $\sqrt{s}=8$ TeV, at both NLL and NNLL for different non-perturbative parameters. A significant part of the uncertainties originates in the variation of the rapidity scale, and the bands at NLL overlap with the NNLL bands. If we compare the two panels, we see that the impact of the non-perturbative contribution is rather small and induce changes to the distributions of similar size as the uncertainty on the prediction at NNLL. The Higgs boson transverse momentum distribution is thus not very sensitive to those parameters. The same conclusion was drawn in \cite{Becher:2012yn}, where a Gaussian model was used to parametrize the non-perturbative contributions. The rightmost panel of Figure~\ref{fig:xsection2}  shows the predictions for the distribution at $\sqrt{s}=13$ TeV, for different values of the non-perturbative parameters. The cross section is larger than at $\sqrt{s}=8$ TeV, but the same conclusions hold regarding the sensitivity to the non-perturbative parameters. It therefore seems unlikely that experimental measurements of the Higgs $q_T$ distribution at the LHC will be precise enough to fix the non-perturbative parameters of gluon TMDs, apart from excluding the most vivid parameter values.
\subsection*{Acknowledgements}
We acknowledge financial support from the European Community under the "Ideas" program QWORK (contract 320389).


\begin{thebibliography}{99}

\bibitem{Collins:2011zzd}
  J.~Collins,
  (Cambridge monographs on particle physics, nuclear physics and cosmology. 32)


\bibitem{GarciaEchevarria:2011rb}
  M.~G.~Echevarria, A.~Idilbi and I.~Scimemi,
  JHEP {\bf 1207} (2012) 002
  [arXiv:1111.4996 [hep-ph]].


\bibitem{Echevarria:2012js}
  M.~G.~Echevarria, A.~Idilbi and I.~Scimemi,
  Phys.\ Lett.\ B {\bf 726} (2013) 795
  [arXiv:1211.1947 [hep-ph]].


\bibitem{Echevarria:2015uaa}
  M.~G.~Echevarria, T.~Kasemets, P.~J.~Mulders and C.~Pisano,
  arXiv:1502.05354 [hep-ph].


\bibitem{Sun:2011iw}
  P.~Sun, B.~W.~Xiao and F.~Yuan,
  Phys.\ Rev.\ D {\bf 84} (2011) 094005
  [arXiv:1109.1354 [hep-ph]].


\bibitem{Catani:2011kr}
  S.~Catani and M.~Grazzini,
  Eur.\ Phys.\ J.\ C {\bf 72} (2012) 2013
   [Eur.\ Phys.\ J.\ C {\bf 72} (2012) 2132]
  [arXiv:1106.4652 [hep-ph]].


\bibitem{Chiu:2012ir}
  J.~Y.~Chiu, A.~Jain, D.~Neill and I.~Z.~Rothstein,
  JHEP {\bf 1205} (2012) 084
  [arXiv:1202.0814 [hep-ph]].


\bibitem{Mantry:2009qz}
  S.~Mantry and F.~Petriello,
  Phys.\ Rev.\ D {\bf 81} (2010) 093007
  [arXiv:0911.4135 [hep-ph]].


\bibitem{Becher:2012yn}
  T.~Becher, M.~Neubert and D.~Wilhelm,
  JHEP {\bf 1305} (2013) 110
  [arXiv:1212.2621 [hep-ph]].


\bibitem{Neill:2015roa}
  D.~Neill, I.~Z.~Rothstein and V.~Vaidya,
  arXiv:1503.00005 [hep-ph].


\bibitem{Ji:2005nu}
  X.~d.~Ji, J.~P.~Ma and F.~Yuan,
  JHEP {\bf 0507} (2005) 020
  [hep-ph/0503015].


\bibitem{Zhu:2013yxa}
  R.~Zhu, P.~Sun and F.~Yuan,
  Phys.\ Lett.\ B {\bf 727} (2013) 474
  [arXiv:1309.0780 [hep-ph]].


\bibitem{Mulders:2000sh}
  P.~J.~Mulders and J.~Rodrigues,
  Phys.\ Rev.\ D {\bf 63} (2001) 094021
  [hep-ph/0009343].


\bibitem{Catani:2010pd}
  S.~Catani and M.~Grazzini,
  Nucl.\ Phys.\ B {\bf 845} (2011) 297
  [arXiv:1011.3918 [hep-ph]].


\bibitem{Boer:2014tka}
  D.~Boer and W.~J.~den Dunnen,
  Nucl.\ Phys.\ B {\bf 886} (2014) 421
  [arXiv:1404.6753 [hep-ph]].


\bibitem{Chiu:2011qc}
  J.~y.~Chiu, A.~Jain, D.~Neill and I.~Z.~Rothstein,
  Phys.\ Rev.\ Lett.\  {\bf 108} (2012) 151601
  [arXiv:1104.0881 [hep-ph]].


\bibitem{Collins:2012uy}
  J.~C.~Collins and T.~C.~Rogers,
  Phys.\ Rev.\ D {\bf 87} (2013) 3,  034018
  [arXiv:1210.2100 [hep-ph]].


\bibitem{Boer:2011kf}
  D.~Boer, W.~J.~den Dunnen, C.~Pisano, M.~Schlegel and W.~Vogelsang,
  Phys.\ Rev.\ Lett.\  {\bf 108} (2012) 032002
  [arXiv:1109.1444 [hep-ph]].


\bibitem{Boer:2013fca}
  D.~Boer, W.~J.~den Dunnen, C.~Pisano and M.~Schlegel,
  Phys.\ Rev.\ Lett.\  {\bf 111} (2013) 3,  032002
  [arXiv:1304.2654 [hep-ph]].


\bibitem{Pisano:2013cya}
  C.~Pisano, D.~Boer, S.~J.~Brodsky, M.~G.~A.~Buffing and P.~J.~Mulders,
  JHEP {\bf 1310} (2013) 024
  [arXiv:1307.3417 [hep-ph]].


\bibitem{Dunnen:2014eta}
  W.~J.~den Dunnen, J.~P.~Lansberg, C.~Pisano and M.~Schlegel,
  Phys.\ Rev.\ Lett.\  {\bf 112} (2014) 212001
  [arXiv:1401.7611 [hep-ph]].


\bibitem{Echevarria:2012pw}
  M.~G.~Echevarria, A.~Idilbi, A.~Sch\"{a}fer and I.~Scimemi,
  Eur.\ Phys.\ J.\ C {\bf 73} (2013) 12,  2636
  [arXiv:1208.1281 [hep-ph]].


\bibitem{Qiu:2000hf}
  J.~w.~Qiu and X.~f.~Zhang,
  Phys.\ Rev.\ D {\bf 63} (2001) 114011
  [hep-ph/0012348].
  
 \end{thebibliography}
\end{document}